\documentclass[reprint,showpacs,aip,apl]{revtex4-1}

\usepackage{graphicx}
\usepackage{color}

\newcommand{\new}[1]{\textcolor{black}{ #1}}

\begin{document}

\title{Thermally induced error: density limit for magnetic data storage}

\author{R. F. L. Evans}
\affiliation{Department of Physics, The University of York, York, YO10 5DD, UK}
\author{R. W. Chantrell}
\affiliation{Department of Physics, The University of York, York, YO10 5DD, UK}
\author{U. Nowak}
\affiliation{Fachbereich Physik, Universit\"at Konstanz, 78457 Konstanz, Germany}
\author{A. Lyberatos}
\affiliation{Department of Materials Science, University of Crete, Heraklion 71003,Greece}
\author{H-J. Richter}
\affiliation{Hitachi Global Storage Technologies, 3403 Yerba Buena Road, San Jose, CA 95135}

\begin{abstract}
Magnetic data storage is pervasive in the preservation of digital
information and the rapid pace of computer development requires ever more
capacity. Increasing the storage density for magnetic hard disk drives
requires a reduced bit size, previously thought to be limited by the thermal
stability of the constituent magnetic grains. The limiting storage density
in magnetic recording is investigated treating the writing of bits as a
thermodynamic process. A 'thermal writability' factor is introduced and it
is shown that storage densities will be limited to 15 to 20 TBit/in$^2$ unless technology can move beyond the currently available write field magnitudes.
\end{abstract}

\pacs{85.70.-w,85.70.Li,75.50.Ss,75.75+a}

\date{\today}

\maketitle

As a technology, magnetic recording has been in existence since the invention of magnetic tape recording in the 1920s and 1930s.  Since the early 1980's, and the introduction of metallic thin film recording media, the industry has seen a rapid increase in storage density;  up to the TByte storage available in today's PC hard drives. Because technology has kept pace with demand, magnetic information storage is now ubiquitous. Having been around for some 60 years, magnetic recording is running into difficulties imposed by physical limitations.

A previous study of the possible limits of recording density was made by Charap et al \cite{charap} . This study predicted an upper limit of 36 Gb/in$^2$ and, remarkably, current technology has already achieved densities over one order of magnitude beyond this value. The reason for this lies in advances in the 'non-magnetic' aspects of recording technology, including error detection and correction and the mechanical actuator systems used to position the read and write sensors, which were not anticipated by the authors of Ref. \onlinecite{charap}. The question is; does there exist a physical upper limit to recording density which cannot be exceeded by improved technology? Here we argue that the limitation is essentially determined by the maximum tolerable Bit Error Rate and certain materials parameters which, critically, includes the saturation magnetisation of the recording medium.

Magnetic recording relies on the storage of information on media comprised of grains of a material with a high magnetocrystalline anisotropy. The grains can be considered as bistable systems capable of representing bits of information in terms of the polarity of the grains. Stability of the information is provided by an anisotropy energy barrier $KV$ where $K$ is the anisotropy constant and $V$ the grain volume. It has long been realised that the phenomenon of 'superparamagnetism' (SPM) defines the upper limit thermal stability of magnetic materials\cite {bl}. In the case of magnetic recording information should be stable for at least 10 years, which leads to an established criterion of $KV/kT>60$ for media design.

Future advances in magnetic recording density will have to circumvent the magnetic recording trilemma\cite{hjr}. The key component of the trilemma is the necessary reduction in grain size for signal to noise reasons. For thermal stability the anisotropy energy must therefore increase, to a point where conventional recording heads are unable to write the medium. Heat Assisted Magnetic Recording (HAMR) \cite{rottmayer,TMcD} is a potential mechanism to solve this problem, by heating the material to the vicinity of its Curie point, where the anisotropy is low, writing the data, and then cooling back to the storage temperature of the medium. In the following we show that under this scenario the fact that the applied field is greater than the coercivity of the medium is an insufficient criterion: thermal fluctuations in the material itself lead to write errors. In fact this is a general problem for all small magnetic elements, ultimately that the recording process must be thermodynamically favourable to be reliable, and so it is the \textit{thermal writability} that determines the ultimate limits of magnetic recording. \new{We specifically address the problem of the ultimate recording medium, which is likely to consist of a combination of Bit Patterned Media and HAMR, although the underlying physics is equally applicable to normal HAMR.}

We proceed by calculating the Bit Error Rate (BER) induced by thermal fluctuations during the write process. Consider the ultimate recording system in which one magnetic grain is sufficient to store a binary '1' or '0'. Our approach is to consider the equilibrium magnetization $m_e$ in the recording context. In conventional recording there are a number of grains per bit so $m_e$ has the meaning of an ensemble average magnetisation. Here, values of $m_e$ less than unity represent the probability of a non-reversed grain in a bit, which gives rise to a dc noise. Consider now the situation of recording one bit of information per grain. Since we are now dealing with individual grains, $m_e$ must be interpreted differently; in terms of \textit{ the probability $p_{sw}$ that the magnetisation is switched into the correct state by the field during the attempt to write the information,} specifically, $p_{sw}=(m_e+1)/2$. Since the Bit Error Rate (BER) is essentially the probability of wrongly recording the bit, we have simply that $BER=1-p_{sw}=(1-m_e)/2$. Considering for simplicity the case of a system with perfectly aligned easy anisotropy axes, using a master equation approach it can be shown that the thermal equilibrium magnetisation is given by
\begin {equation}\label{me}
m_e=\tanh \left (\frac{\mu \mu_0 H}{kT} \right )
\end {equation}
where $\mu = M_s V$ is the magnetic moment of the grain with $M_s$ the material saturation magnetization and $V$ the particle volume. \new{Eq. \ref{me} represents the ensemble-averaged magnetisation of a collection of grains or the time-average magnetization of a single grain.} The physical importance of Eq.~\ref{me} is that, if the system is left for an infinite time in an applied field, the best case is to achieve the thermal equilibrium value for the magnetization. Consequently for low values of $\frac{\mu \mu_0 H}{kT}$ the equilibrium magnetization can be much less than 1, which is realised for small particle volumes or elevated temperatures. By rearrangement of the equivalent exponential form of Eq.~\ref{me} and applying the relevant limit of $\frac{\mu \mu_0 H}{kT}>>1$ we have the result that
\begin {equation}
BER= \frac{1-m_e}{2} = \exp\left (-\frac{2\mu \mu_0 H_{wr}}{kT} \right )
\label{eq:bera_main}
\end {equation}
where $H_{wr}$ is the field available from the write transducer. \new{Physically, Eq. \ref{eq:bera_main} represents the fact that, given sufficient energy to overcome the energy barrier, the energy difference between the parallel and anti-parallel states  depends only on $H_{wr}$ and is independent of anisotropy. } Eq. \ref{eq:bera_main} introduces a new factor in the writability of stored information. The writability is normally considered in terms of the field needed to switch the magnetisation. However, it is clearly necessary, in addition, to maintain a large value of $\mu \mu_0 H_{wr}/kT$ in order to avoid thermally driven switching failures and to achieve the required BER; we define this as the {\em thermal writability} of the medium. \new{We note that large values of moment are required in order to maintain a low anisotropy field to assist switching. However, the thermal writability is a new requirement beyond that of achieving switching within the usual trilemma.}

\begin{figure}[htb]
  \begin{center}
  \includegraphics[width=8.5cm]{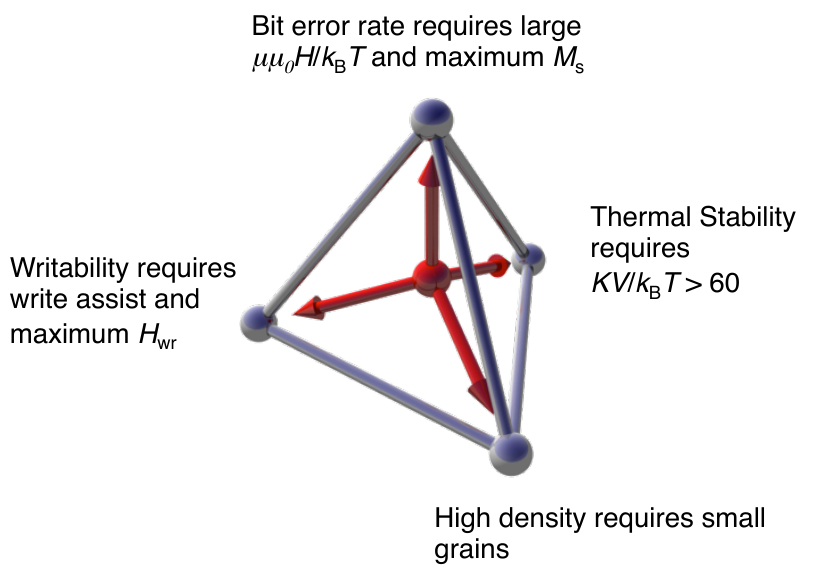}\\
  \end{center}
 \caption{Schematic of the 'quadrilemma' of magnetic recording. The decrease of grain volume requires an increase in the anisotropy constant $K$ for thermal stability and also maximisation of the saturation magnetisation $M_s$ to ensure thermal writability.}
 \label{fig:quadrilemma}
\end{figure}

Hence the design of ultra-high density recording techniques is a {\em quadrilemma}, shown schematically in Fig.~\ref{fig:quadrilemma}, rather than a trilemma which underlies current media and system design. In addition to decreasing the volume of the bits and increasing $K$ (thereby leading to the requirements for write assist) the value of $M_s$ must remain high in order to maintain writability as the grain size decreases. A further factor is the nature of the role of the write field. It is known to be important to keep the write field as large as possible, and generally speaking this is considered to result from the need to lower the energy barriers sufficiently to switch the magnetisation. However, our model suggests that increasingly large fields are important in maintaining a tolerable BER because of the requirement for thermal writability.

\begin{figure}[htb]
\center
\includegraphics[width=8.5cm]{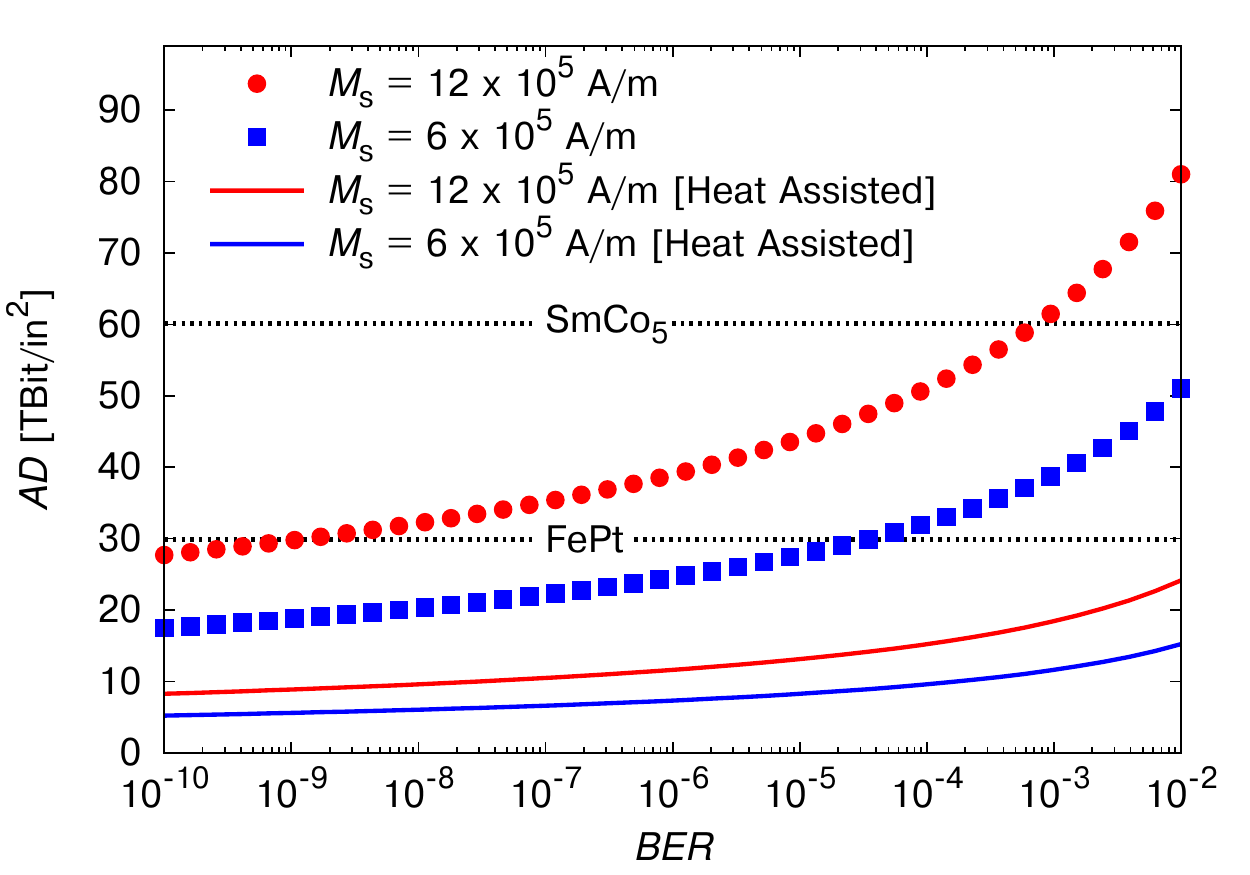}
\caption{Calculated areal density ($AD$) as a function of the tolerable Bit Error Rate (BER). Symbols show the results for the non-heat assisted case and the solid lines are calculated under the assumption of a heat-assist with a write temperature $T_{wr}=740$K and a Curie temperature $T_c=750$K. Calculations are given for values of the saturation magnetisation of 12 and 6 $\times 10^5$ A/m at 0K. The maximum (stability limited) values of areal density ($AD_{ts}$) are shown for FePt and SmCo$_5$, with anisotropies of $1 \times 10^7$ and $1 \times 10^8$ J/m$^3$ respectively. For a given BER the achievable value of density is the minimum of $AD$ and $AD_{ts}$. }
\label{fig:Hwrite10kOe}
\end{figure}

In order to evaluate the likely achievable areal densities we proceed from Eq. \ref{eq:bera_main}. Solving for the volume and assuming that the Areal Density ($AD$) is given approximately by $V^{-2/3}\alpha$, where $\alpha = 0.5$ is the areal packing fraction of the storage islands, it is straightforward to show that
\begin {equation}
AD= \left (  \frac{2 M_s \mu_0 H_{wr}}{kT \ln \left [(BER)^{-1} \right ]} \right )^{2/3} \alpha.
\label{eq:AD}
\end {equation}
Note that the areal density is now determined by two conditions. Firstly it cannot be larger than determined by the thermal stability criterion\cite{weller} $KV/kT>60$. Using the approximation $AD= V^{-2/3}\alpha$ this gives

\begin {equation}
AD_{ts}= \left (\frac{K}{60 kT } \right )^{2/3} \alpha.
\label{eq:ADts}
\end {equation}
Secondly, the $AD$ must also be less than the value determined by the allowed BER, given by Eq. \ref{eq:AD}. Consequently the achievable areal density is the minimum of $AD_{ts}$ and the value determined from Eq. \ref{eq:AD}.

As an illustrative example, Fig.~\ref{fig:Hwrite10kOe} shows calculation of Areal Densities for the (reasonable) case of a write field $\mu_0 H_{wr}=1 $T. Values of $AD_{ts}$ are shown for FePt and SmCo$_5$, two high-$K$ materials. Consider first the case of no heat assist. The conclusion from these calculations is that very large areal densities are possible, but that the BER needed to achieve maximum density increases with decreasing $M_s$, which becomes an important factor in the quadrilemma now seen to govern media and system design. Taking first FePt, with a value of $M_s=12 \times 10^5$ A/m the $AD_{ts}$ value is achieved with a BER of $10^{-9}$, consistent with today's technology. However, a for a value of $M_s=6 \times 10^5$ A/m the $AD_{ts}$ value requires a significant increase of BER to around $3\times 10^{-5}$. Increasing the anisotropy value imposes even more stringent requirements on the BER. Here calculations are given to a limiting value of BER=$10^{-2}$, which would present an enormous technical challenge. For the $AD_{ts}$ value corresponding to SmCo$_5$, achieving the maximum areal density requires a BER value of $7 \times 10^{-4}$  for $M_s=12 \times 10^5$ A/m. For $M_s=6 \times 10^5$ A/m $AD_{ts}$ cannot be reached; the areal density is limited by the BER (assuming that the highly challenging figure of $10^{-2}$ is achievable) to about 51 TBit/in$^2$.

These figures suggest that extremely high densities are possible with magnetic recording. However, if one takes account of the fact that heat assist is expected to be necessary to write on high anisotropy media a different picture appears. To extend the calculations to HAMR we introduce the temperature dependence of $M$ using the approximation proposed by Arrott \cite{arrott}
\begin{equation}
M_s(T) = M_s(T=0)(1-(T/T_c)^2)^{1/2},
\end{equation}
with $T_c$ the Curie temperature of the material,  giving
\begin {equation}
AD_{ha}= \left ( \frac{2 M_s(T_{wr}) \mu_0 H_{wr}}{kT_{wr} \ln \left [(BER)^{-1} \right ]} \right )^{2/3} \alpha ,
\label{eq:ADhamr}
\end {equation}
 with $T_{wr}$ the writing temperature, as the BER limited areal density for heat assisted recording. In Fig.~\ref{fig:Hwrite10kOe} results are included assuming a heat assist with a temperature of $T_{wr} = 740$K and $T_c=750$K. It can be seen that there is a dramatic reduction in the BER limited areal density, and that the thermal stability limit, even for FePt is not reached. The results for the case with heat assist are, of course, the most realistic since some form of write-assist is necessary for recording on high anisotropy media and heat assistance is seen as the most likely solution. The important finding of the current work is that heat assist transforms the areal density 60$kT$ from being thermal stability limited to BER limited, leading to much lower limiting values.

Our findings indicate that the reduction in \emph{grain} volume has not one but two important consequences. The first of these is the thermal stability requirement which leads to the necessity of large $K$ values and the resultant problems with writability and the well-known 'trilemma' for media design. However, the reduction in grain volume also lowers the value of $\mu \mu_0 H_{wr}/kT$, which must be compensated for by increasing the value of the saturation magnetisation $M_s$. Consequently recording media design in fact becomes a 'quadrilemma' (see Fig. \ref{fig:quadrilemma}). Thus the future of magnetic recording requires moving to a new paradigm in which the thermal writability introduced here is treated on an equal footing to the conventional writability, which is essentially the requirement of a sufficiently large field to switch the magnetisation state.

In summary, we present arguments which suggest that the areal density of magnetic recording is likely to be limited by the achievable BER due to purely thermodynamic effects. The theoretical approach suggests that the optimisation of recording density must be considered as a {\em quadrilemma} including the requirement of maintaining a saturation magnetisation sufficiently large to ensure writability. Our calculations suggest that between 15 and 20 TBit/in$^2$ may be achievable with heat assisted writing. \new{It should be noted that the calculations are sensitive to the temperature to which the system is heated. Consequently the thermal writability should be considered as an important factor in the optimization of HAMR. We also stress that the calculations given here are strictly applicable for the ultimate recording system comprised of Bit Patterned media with HAMR. However, the underlying physics of thermal writability is also a limiting factor in HAMR although its formulation is more complex.}

Finally, we note that the calculations given here are made under the assumption of a write field of 1T, which is realistic for the inductive technology used in today's write transducers being limited by saturation magnetization.  However, it is clear from the expressions given here that a significant increase in write field would greatly extend magnetic recording technology. The  discovery of optomagnetic reversal \cite{stanciu07,vahaplar09}, which generates effective fields significantly higher that inductive technology, suggests that novel approaches may be possible. Certainly some means of generating increased writing fields is necessary in order to overcome the thermal writability problem as densities increase.

The authors gratefully acknowledge the support of the European Community's Seventh Framework Programme (FP7/2007-2013) under grant agreements NMP3-SL-2008-214469 (UltraMagnetron) and N 214810 (FANTOMAS).

\end{document}